\documentclass[preprint,12pt]{aastex631}  
\usepackage{epsfig}
\usepackage{soul}
\usepackage{color}

\usepackage{physymb}
\usepackage{mathrsfs}

\def\oviii{O\,{\sc viii]}}
\def\ovii{O\,{\sc vii]}}

\def\fexxv{Fe\,{\sc xxv}}
\def\fexxvi{Fe\,{\sc xxvi}}

\def\siiv{Si\,{\sc iv}}

\def\nex{Ne\,{\sc x}}

\def\cii{C\,{\sc ii}}

\def\civ{C\,{\sc iv}}

\def\ovii{O\,{\sc vii}}
\def\oviii{O\,{\sc viii}}

\def\civ{C\,{\sc iv}}

\def\ovii{O\,{\sc vii}}
\def\nv{N\,{\sc v}}

\def\mathv{\textbf{\em v}}

\def\cm{\ifmmode {\rm cm}^{-1} \else cm$^{-1}$ \fi}
\def\s{\ifmmode {\rm s}^{-1} \else s$^{-1}$ \fi}
\def\cc{\ifmmode {\rm cm}^{-3} \else cm$^{-3}$ \fi}
\def\cs{\ifmmode {\rm cm}^{-2} \else cm$^{-2}$ \fi}
\def\g{\ifmmode \gamma \else $\gamma$\fi}
\def\G{\ifmmode \Gamma \else $\Gamma$\fi}
\def\Gs{\ifmmode \Gamma~ \else $\Gamma~$\fi}

\def\gc{\ifmmode \gamma_{\rm c} \else $\gamma_{\rm c}$ \fi}

\def\gsim{\mathrel{\raise.5ex\hbox{$>$}\mkern-14mu
             \lower0.6ex\hbox{$\sim$}}}
\def\lsim{\mathrel{\raise.3ex\hbox{$<$}\mkern-14mu
             \lower0.6ex\hbox{$\sim$}}}
\def\simless{\mathbin{\lower 3pt\hbox
     {$\rlap{\raise 5pt\hbox{$\char'074$}}\mathchar"7218$}}}   
\def\simmore{\mathbin{\lower 3pt\hbox
     {$\rlap{\raise 5pt\hbox{$\char'076$}}\mathchar"7218$}}}   
\def\Msun{M_\odot}                                
\def\deg{^\circ}
\def\aa{\buildrel_{\circ}\over{\mathrm{A}}}

\newcommand{\Alfven}{Alfv$\acute{\rm e}$n~}

\def\gro1655{GRO~J1655-40}
\def\4u1630{4U1630-472}
\def\h1743{H1743-322}
\def\grs1915{GRS1915+105}

\def\xmm{{\it XMM-Newton}}
\def\chandra{{\it Chandra}}
\def\nustar{{\it NuSTAR}}
\def\swift{{\it Swift}}
\def\hst{{\it HST}}

\shorttitle{X-ray Obscuration of AGNs with Disk-Wind}

\shortauthors{Fukumura et al. 2024}

\begin{document}


\title{Dual Role of Accretion Disk Winds as X-ray Obscurers and UV Line Absorbers in AGN}


\date{\today}

\author[0000-0001-5709-7606]{Keigo Fukumura}
\affiliation{Department of Physics and Astronomy, James Madison University,
Harrisonburg, VA 22807; fukumukx@jmu.edu}

\author[0000-0002-4992-4664]{Missagh Mehdipour}
\affiliation{Space Telescope Science Institute, 3700 San Martin Drive, Baltimore, MD 21218}

\author{Ehud Behar}
\affiliation{Physics Department, Technion, Haifa 32000, Israel}

\author{Chris Shrader}
\affiliation{Astrophysics Science Division, NASA/Goddard Space Flight Center, Greenbelt, MD 20771}
\affiliation{Department of Physics, Catholic University of America, Washington, DC 20064}

\author{Mauro Dadina}
\affiliation{INAF, Osservatorio di Astrofisica e Scienza dello Spazio di Bologna, via P. Gobetti 93/3, 40129 Bologna, Italy}

\author{Demosthenes Kazanas}
\affiliation{Astrophysics Science Division, NASA/Goddard Space Flight Center, Greenbelt, MD 20771}

\author{Stefano Marchesi}
\affiliation{Dipartimento di Fisica e Astronomia, Universit\`{a} degli Studi di Bologna, via Gobetti 93/2, 40129 Bologna, Italy}
\affiliation{Department of Physics and Astronomy, Clemson University, Kinard Laboratory of Physics, 140 Delta Epsilon Ct., Clemson, SC 29634}
\affiliation{INAF, Osservatorio di Astrofisica e Scienza dello Spazio di Bologna, via P. Gobetti 93/3, 40129 Bologna, Italy}

\author{Francesco Tombesi}
\affiliation{Astrophysics Science Division, NASA/Goddard Space Flight Center, Greenbelt, MD 20771}
\affiliation{Department of Astronomy, University of Maryland, College Park, MD20742}
\affiliation{Department of Physics, University of Rome ``Tor Vergata", Via della Ricerca Scientifica 1, I-00133 Rome, Italy}
\affiliation{INAF Astronomical Observatory of Rome, Via Frascati 33, 00078 Monteporzio Catone (Rome), Italy}

\begin{abstract}
\baselineskip=15pt

X-ray obscuration of active galactic nuclei (AGNs) is considered in the context of ionized winds of stratified structure  launched from  accretion disks. We argue that  a Compton-thick layer of a large-scale disk wind can  obscure continuum X-rays and also lead to broad UV  absorption such as in the blue wing of \civ; the former originates from the inner wind while the latter from the outer wind as a dual role.      
Motivated by a number of observational evidence showing strong AGN obscuration phenomena in Seyfert 1 AGNs {\bf such as NGC~5548}, we demonstrate in this work, by utilizing a physically-motivated wind model coupled to post-process radiative transfer calculations, that an extended disk wind under certain physical conditions (e.g. morphology and density) could naturally cause a sufficient obscuration qualitatively consistent with UV/X-ray observations. Predicted UV/X-ray correlation is also presented as a consequence of variable spatial size of the wind in this scenario.   

\end{abstract}





\baselineskip=15pt

\section{Introduction}

Outflows are among the most fundamental features almost ubiquitously seen in  active galactic nuclei (AGNs) being manifested in UV/X-ray spectra in the form of blueshifted absorption lines \citep[e.g.][]{CKG03, Tombesi10,Arav15}. Galactic outflows beyond parsec-scale in the host galaxies have also been seen in the form of molecular winds, most likely having a physical origin conncted to the hotter and faster innermost  AGN winds \citep[e.g.][]{Faucher-Giguere12, Wagner13, Tombesi15}. In particular, ionized outflows seen in X-ray, 
conventionally termed as warm absorbers \citep[e.g.][]{Blustin05, Steenbrugge05, McKernan07,Laha14, Laha16}, 
may  be related to accretion disks, narrow line region or torus around supermassive black holes (BHs), although a definitive identification of their launching mechanisms remains elusive to date (i.e. thermal, radiation and magnetic). These accretion-disk winds of a line-of-sight velocity $v$ consist of a variety of chemical elements at different ionization states. Particularly in X-rays, most notable is hydrogen/helium-like absorbers through resonance transition (e.g. \ovii/\oviii, \nex\ and \fexxv/\fexxvi) independently characterized by hydrogen-equivalent column density $N_H$ and ionization parameter $\xi$. 

In addition to the observed ionized outflows, a number of AGNs are further known to undergo significant spectral variations through UV absorption and soft X-ray obscuration by  variable optically-thick materials. 
In some case, these are suggestive of localized obscuration by  Compton-thick media conceived in the innermost circumnuclear region for Seyfert 2 AGNs \citep[e.g.][]{Marchesi18,Marchesi22}. While not clear how and where the implied obscurers are produced and driven, their origin  appears to be closely related to large-scale disk winds perhaps extending all the way out to  broad line region (BLR; 
e.g. \citealt{Kaastra14}; \citealt{M22}, hereafter, M22; \citealt{Kara21}; \citealt{Homayouni23}). These X-ray obscurers  have been extensively studied and analyzed in conjunction with the response of their UV counterpart found in an increasing fraction of Seyfert 1 AGNs to date; e.g. NGC~3783 \citep[e.g.][]{M17,Kriss19}, NGC~5548 (e.g. \citealt{Kaastra14}; \citealt{M15}; \citealt{Dehghanian21}; M22), Mrk~817 \citep[e.g.][]{Cackett23,Partington23,Homayouni23} and NGC~985 \citep[e.g.][]{Ebrero16}, among others, based primarily on synergistic monitoring programs such as {\tt AGN STORM} collaboration \citep[][]{Dalla-Bonta16}.

For example, the 2013 campaign on NGC~5548 (Kaastra et al. 2014) showed that the X-ray obscurer in NGC 5548 consists of two different ionized components. The 1st component with $N_H \sim 1.2 \times 10^{22}$ cm$^{-2}$ covers about 90\% of the X-ray source, while the 2nd component has higher column density ($N_H \sim 10^{23}$ cm$^{-2}$) and lower covering fraction (30\%). This 2nd component with lower ionization parameter is thought to be denser clumps embedded in the medium of the 1st component. Further photoionization modeling studies of the obscurer in NGC~5548 (Kriss et al. 2019) found that it is very challenging to explain both the X-ray obscuration and the broad UV absorption lines together consistently. Most likely because of the complex and multi-component nature of the obscuring wind, finding a unique photoionization solution is not feasible, and in particular the ionization parameter of the obscurer remains uncertain. This complexity makes it difficult to derive other physical parameters of the obscurer. Nonetheless, the broad UV absorption lines (such as \civ, \nv, \siiv, and Ly$\alpha$) show the outflow velocity of the obscurer is dominant at around 2,000 km/s , reaching up to 5,000 km/s (Kaastra et al. 2014). The previous X-ray and UV studies have found that the obscurer varies from short timescales (ks/days; Kaastra et al. 2014, Di Gesu et al. 2015, Cappi et al. 2016) to long timescales (month/years; M22). Based on the broad UV absorption lines, which partially cover the broad emission lines, Kaastra et al. (2014) inferred the obscurer is at a distance of 2-7 light days from the nucleus. We refer to these previous studies of NGC~5548 for more details on the X-ray and UV observational characteristics of the obscurer. Any reliable time lag for the obscurer itself is yet to be known to date.

These studies have clearly indicated  that  X-ray obscuration and its UV signatures are physically correlated through variability in terms of certain quantities such as UV/X-ray covering fractions, the strength of \civ\ absorption line and X-ray hardening,  for example.  
Peculiar time-lags are seen in the observed \civ\ light curves based on an intensive reverberation mapping campaign with HST for Mrk~817 indicative of an obscuring cloud dynamically affecting ionization balance in the gas \citep[e.g.][]{Homayouni23}.

As a canonical Seyfert 1 AGN, NGC~5548 ($z=0.01717$) has been extensively studied  with most of the space-based observatories such as \chandra, \xmm, \nustar, \swift, and \hst\ for exhibiting a plethora of rich UV/X-ray features. 
Among others, M22 show that this AGN has gone through an evolutionary phase of X-ray obscuration since 2012, accompanied by broad (e.g. \civ) and narrow (e.g. \cii) UV absorption features
($\sim 1,000-5,000$ km~s$^{-1}$; \citealt{Kaastra14}). While the  column density of the observed X-ray warm absorbers is found to have a typical value of $N_H \sim 10^{22}$ cm$^{-2}$ during the unobscured state, it is found that the observed spectra during a series of obscured epochs (i.e. 2012 onwards) require an additional higher column (i.e. $N_H \sim 10^{23}$ cm$^{-2}$). Detailed spectral analyses suggest that  the observed absorption in UV and X-ray obscuration  can be systematically accounted for by high column with variable covering fractions in UV ($C_{\rm UV} \sim 0-0.3$) and X-ray ($C_X \sim 0.5-0.95$), implying an intriguing scenario  in which the UV absorber nearly vanishes while X-ray obscuration is still significant. On another interesting note, the observed \civ\ absorption appears to be imprinted almost exclusively in the blue tail of the observed \civ\ emission line, while very weakly (if not none) in the red tail.  

Another similar observational signature showing a coexistence of UV absorbers and X-ray obscurers has been reported in NGC~3783 ($z=0.00973$)  with the observed absorption, emission and extinction effects being simultaneously analyzed, resulting in nine distinct X-ray warm absorber components \citep[e.g.][]{Mao19}. Multi-epoch synergistic observations of NGC~3783 have further revealed the presence of X-ray-obscuring column ($N_H \sim 10^{23}$ cm$^{-2}$) in \xmm\ spectra simultaneous with the UV absorbers (e.g. \civ\ line up to $\sim 6,000$ km~s$^{-1}$) in the observed \hst/COS data \citep[][]{Kriss19}.   

These unique case studies clearly point to the importance of a fundamental coupling that underlies the separate phenomena seen in UV and X-ray.  Motivated by these observational facts, it is suggested  that fragmentation of disk-winds into  clumpy clouds due to thermal instabilities  \citep[e.g.][]{Waters22} might be among plausible physical configurations consistent with multi-wavelength data (e.g. M22; \citealt{Homayouni23}). Nonetheless, a definitive characterization of the implied absorbing gas and obscuring materials has yet to be fully investigated from a theoretical standpoint in a systematic fashion. 

In this work, we consider the case of a continuous, stratified wind launched over a large radial extent of an AGN accretion disk  by utilizing the existing theoretical framework of magnetohydrodynamic (MHD) disk wind model (i.e.. \citealt{BP82}; \citealt{CL94})
%
%
of certain stratification in wind properties\footnote[1]{Note, however, that the most essential factor of our wind model in this work is its stratification in density and velocity, for example, and hence the exact driving mechanism is not limited to MHD-driving in this respect. }, which has already been shown to successfully explain diverse wind absorption phenomena; e.g. X-ray outflows (\citealt{F10a}, hereafter F10; \citealt{F17}; \citealt{F18}; \citealt{F21}), ultra-fast outflows \citep[][]{F15,F22}, and broad UV \civ\ features \citep[][]{F10b}.

We adopt the wind model in F10 to simulate various wind morphologies, which are then  coupled to post-process radiative transfer calculations with {\tt xstar} \citep[][]{Kallman01}  to solve for ionization equilibrium of irradiated outflows. As a consequence, an emitted UV/X-ray continua are absorbed and obscured by ionized winds from which we compute synthetic spectra via calculated ionic column densities of \civ\ ions as the radiation penetrates progressively through winds. Our primary focus in the present work is to explore observable correlations (or their absence) between the degree of X-ray obscuration and corresponding UV absorption signature traced via \civ\ line. 

In \S 2, we briefly summarize the basics of MHD wind models from F10. In \S 3, we present our results demonstrating its synergistic concordance to the derived observational facts. We discuss the implications of our model and results in \S 4, followed by Summary and Discussion in \S 5.

\begin{deluxetable}{l||cccccccccc}
\tabletypesize{\small} \tablecaption{Stratified Disk Wind Parameters$^\dagger$} \tablewidth{0pt}
\tablehead{Morphology & $F_o$ & $H_o$  & $(dR/dZ)_o$ &  $\theta_A$ & $\theta_4$   & $n_{\rm in}(30\deg)^\ddagger$ for $n_{12}=2.4$  }
\startdata
%
Wind 1  & 0.15 & -1.58  & 0.98 & $\sim 30\deg$ & $\sim 1\deg$  & $6.0 \times 10^{10}$ cm$^{-3}$  \\
Wind 2  & 0.15 & -1.80  & 1.26 & $\sim 46\deg$ & $\sim 3\deg$   & $4.4 \times 10^9$ cm$^{-3}$  \\ 
Wind 3 & 0.065 & -1.7 & 1.28 & $\sim 38\deg$  & $\sim 6\deg$ & $2.9 \times 10^9$ cm$^{-3}$ \\
Wind 4 & 0.03 & -2.46 & 1.87 & $\sim 53\deg$  & $\sim 28\deg$   & -   \\ 
Wind 5  & 0.05 & -2.9 & 5.3 & $\sim 75\deg$ & $\sim 39\deg$  & -   \\
Wind 6  & 0.05 & -3.15 & 5.7 & $\sim 76\deg$  & $\sim 47\deg$   & - \\
\enddata
\label{tab:tab1}
\vspace*{0.2cm}
$^\dagger$ See the text in \S 2 for the physical significance of the parameters. 
$^\ddagger$ Wind density at its inner edge for $30\deg$ LoS. We assume $p=1.1$ for all the wind solutions. 
\end{deluxetable}

\begin{figure}[t]
\begin{center}
\includegraphics[trim=0in 0in 0in
0in,keepaspectratio=false,width=2.6in,angle=-0,clip=false]{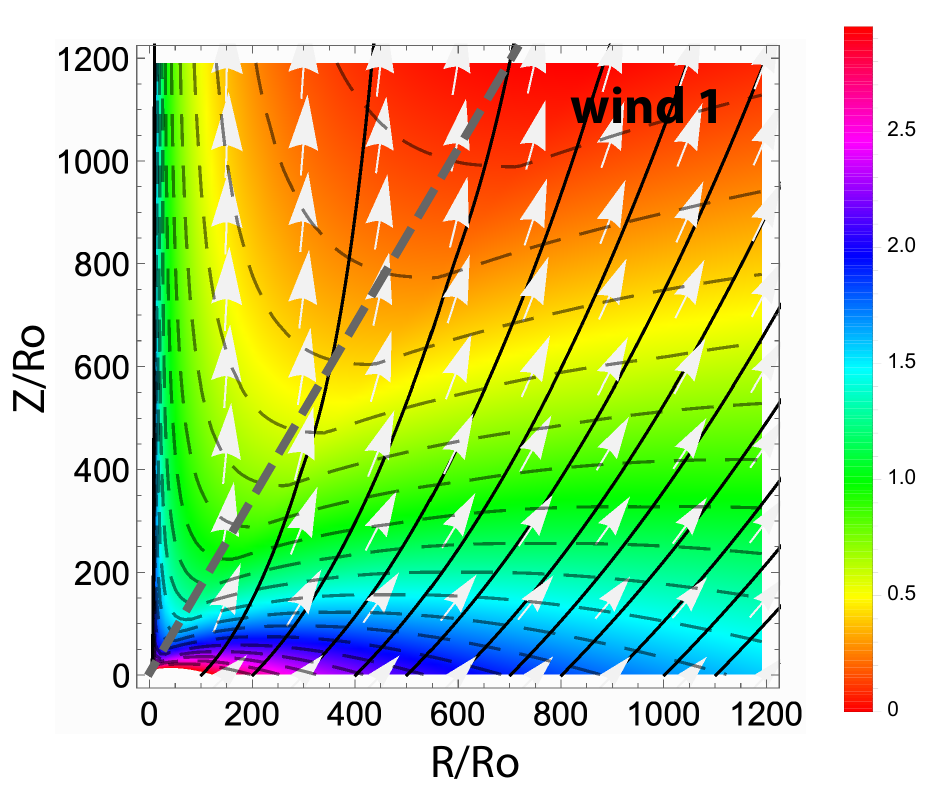}\includegraphics[trim=0in 0in 0in
0in,keepaspectratio=false,width=2.5in,angle=-0,clip=false]{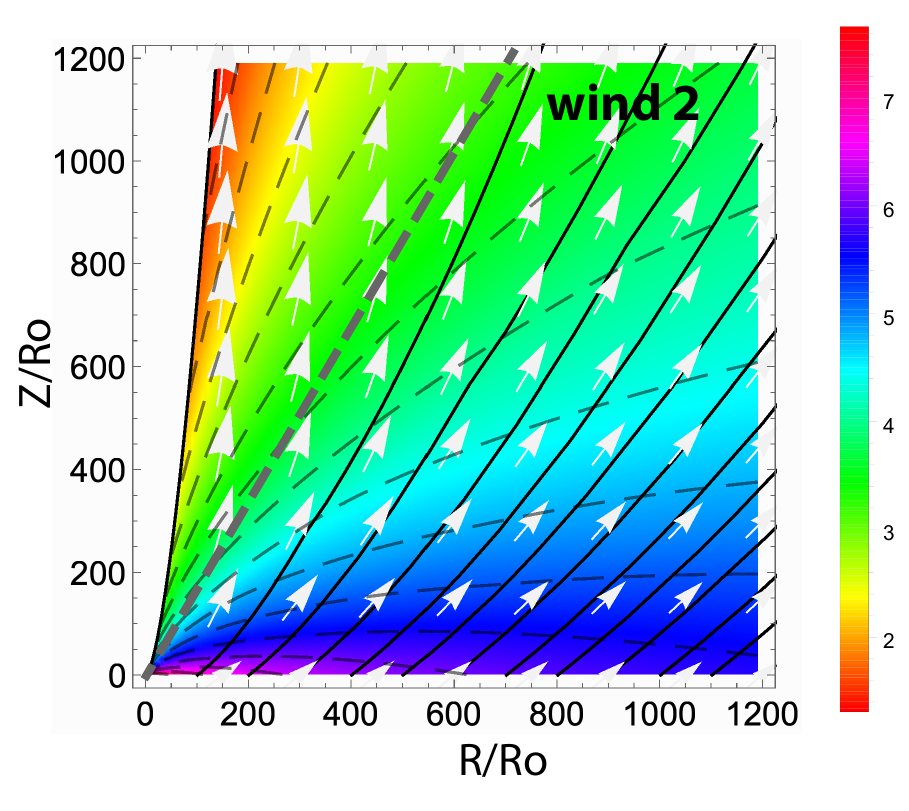}\includegraphics[trim=0in 0in 0in
0in,keepaspectratio=false,width=2.5in,angle=-0,clip=false]{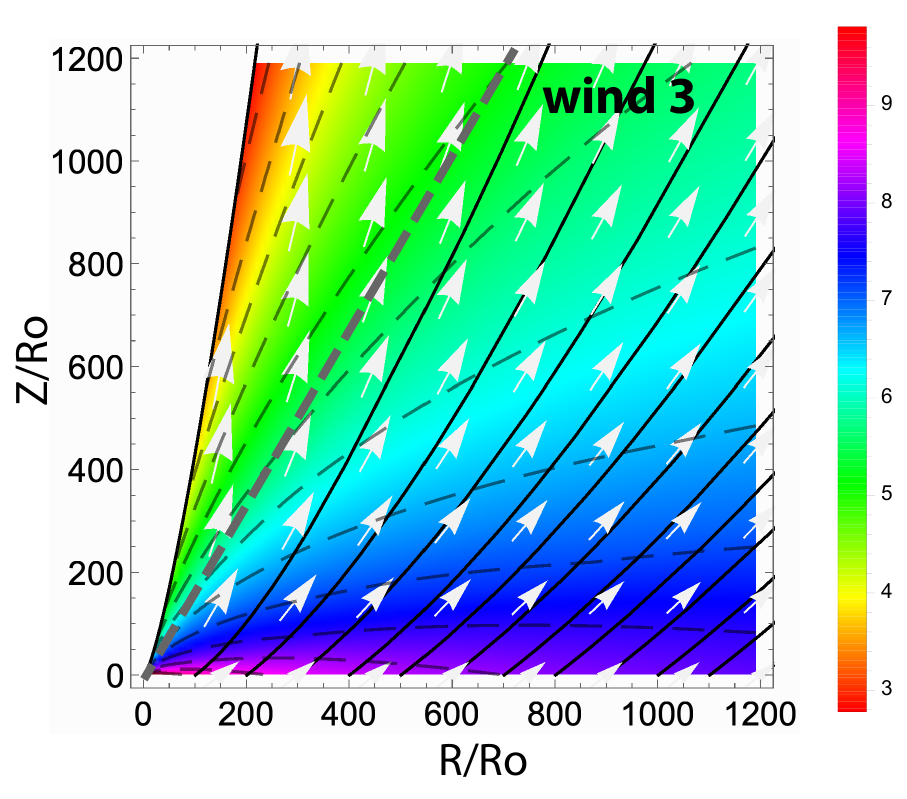} \\
\includegraphics[trim=0in 0in 0in
0in,keepaspectratio=false,width=2.5in,angle=-0,clip=false]{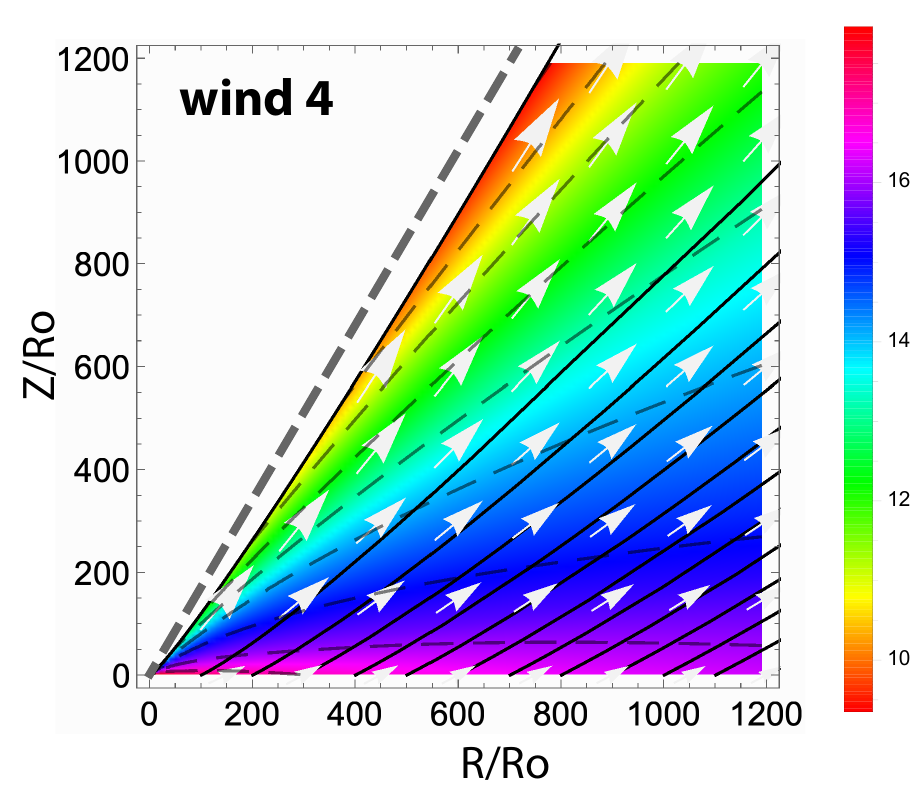}\includegraphics[trim=0in 0in 0in
0in,keepaspectratio=false,width=2.5in,angle=-0,clip=false]{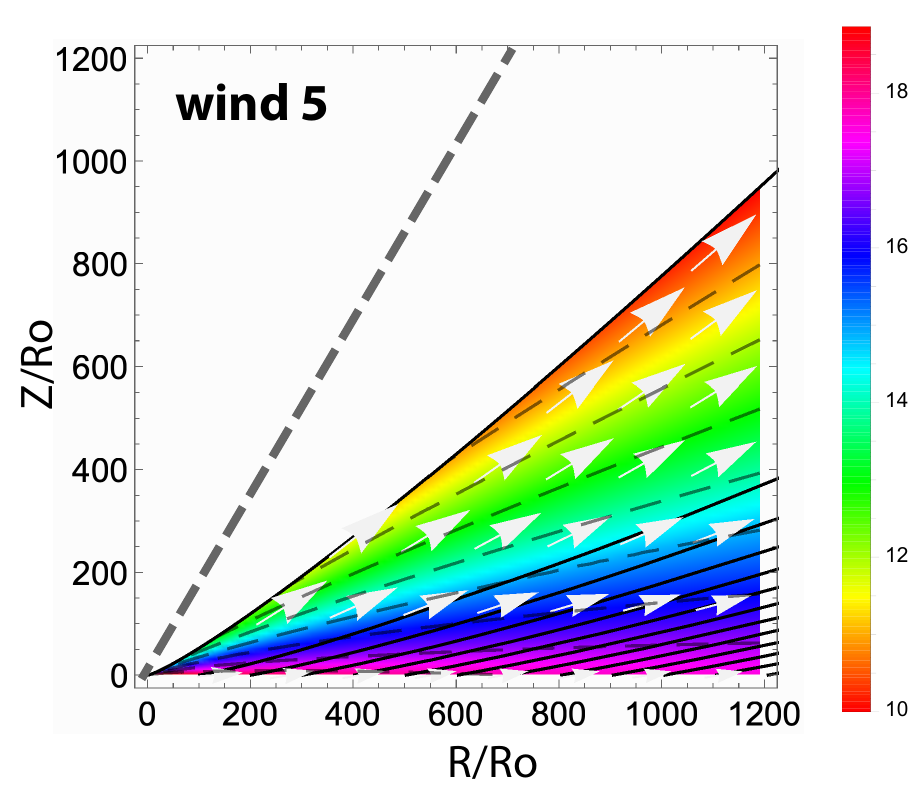}\includegraphics[trim=0in 0in 0in
0in,keepaspectratio=false,width=2.5in,angle=-0,clip=false]{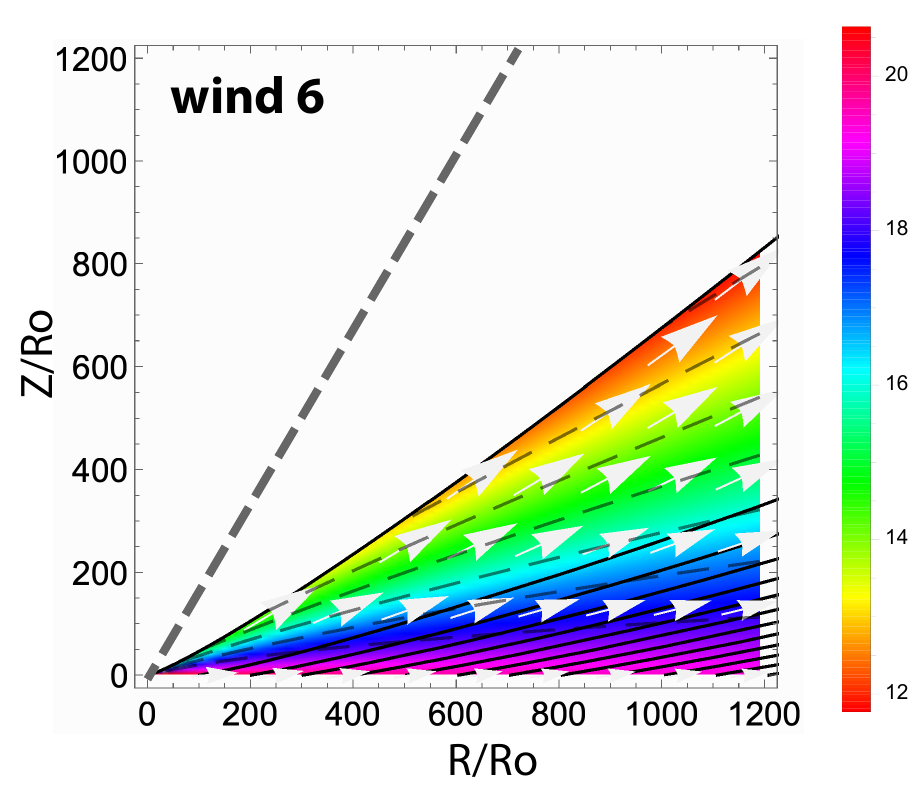}
\end{center}
\caption{Poloidal morphology in $(R, Z)$-coordinates of six simulated MHD-driven disk winds in the innermost region (up to $\sim 10^3 R_o$) for various sets of conserved quantities, as listed in Table~1, showing color-coded normalized density distribution $n(r,\theta)$, density contours (dashed curves), poloidal magnetic field lines (solid curves) and poloidal velocity field (white arrows). Distances are given in units of $R_o \equiv R_{\rm ISCO}$ for $M/\Msun=3 \times 10^7$ as a fiducial black hole mass in NGC~3783 \citep[][]{VestergaardPeterson06} assuming $p=1.1$. LoS of $30\deg$ is given in dashed line. {\bf {\tt Wind 4-6} would produce no obscuration at all for low inclination ($\sim 30\deg$), hence excluded from further discussion in \S3-5. See the explanation in \S 2.} }
\label{fig:wind}
\end{figure}

\section{Basics of the Stratified Wind Model}

As discussed in detail in F10, a classical type of self-similar MHD winds employed in this work \citep[i.e.][]{BP82,CL94} exhibits a characteristic profile \citep[see also][]{K12,Kraemer18,K19,Jacquemin-Ide20}. The most essential part of the model is the radial and angular dependence of the wind density $n(r,\theta)$ and velocity fields $\mathv(r,\theta)=(v_r,v_\theta,v_\phi)$ where 
\begin{eqnarray}
n(r,\theta) = n(r) f(\theta) \equiv n_{\rm 12} \left(\frac{r}{R_{\rm in}} \right)^{-p} f(\theta) ~~~{\rm and} ~~ v_r = v_{\rm LoS}(r,\theta) \sim v_{K} \left(\frac{r}{R_{\rm in}}\right)^{-1/2} g(\theta), \label{eq:eqn2}
\end{eqnarray}
where $n_{\rm 12}$ is the wind density normalization (in units of $10^{12}$ cm$^{-3}$) on the footpoint of the innermost wind streamline anchored at $r=R_{\rm in}$, being typically identified with the innermost stable circular orbit (ISCO), and $v_{K}$ is the Keplerian velocity at $r=R_{\rm in}$. The angular functions, $f(\theta)$ and $g(\theta)$, are numerically calculated by solving  the Grad-Shafranov equation in conjunction with the ideal MHD equations\footnote[2]{Hence, our  framework is NOT force-free or magnetically-arrested \citep[e.g.][]{Tchekhovskoy11}.} (see F10). 
It is reminded that the wind motion along a streamline is smoothly and rapidly converted from toroidal ($v_\phi$) to poloidal ($v_{r,\theta}$) direction across the \Alfven point, which is mainly described by $g(\theta)$. 
On a separate note, the model predicts $v_{\rm LoS} \propto \xi^{1/(4-2p)}$ (F10) in agreement with some AGN X-ray warm absorbers when $p \sim 1.1-1.2$ \citep[e.g.][]{B09,Detmers11,Laha21}. 

To better exploit various wind structures allowed by the model, we first consider six distinct wind morphologies as shown in {\bf Table~1} for different sets of plasma parameters being conserved along a streamline; i.e. particle flux to magnetic flux ($F_o$) and total angular momentum of plasma ($H_o$).  The total plasma  energy of the wind particles, $J_o$,  is roughly conserved as $J_o \simeq -1.5$ in all cases.
The  magnetic field orientation on the foot point of winds (i.e. tangential slope of a field line on the disk surface) is defined by $(dR/dZ)_o \equiv dR/dZ(r=R_o, Z \sim 0)$ in each case. The wind's poloidal geometry is described by opening angle of wind at the \Alfven point  ($\theta_A$) and at the distance of $z/R_{\rm ISCO}=10^4$ ($\theta_4$). 
Detailed definition and description of the model parameters are provided in F10. 
With $n_{12}=2.4$ and $p=1.1$, the innermost wind density along $30\deg$ LoS, $n_{\rm in}(30\degr)$, is computed for {\tt wind 1-3} in Table~1. 
%
In order to {\bf avoid} additional model dependencies,  we make two assumptions; (i) $p=1.1$ not arbitrarily but for the observational implications from AMD analysis as described above. Winds with $p \ll 1$ or $p \gg 1$ would also unrealistically overproduce or underproduce UV absorbers, which would be inconsistent with UV/X-ray data. (ii) $\theta=30\deg$ as often conceived for Seyfert 1 AGNs in the conventional unification paradigm.

%
We note, for example, that the {\tt wind 2} considered here is characterized by $\tau \sim 0.026 \ll 1$ in the innermost layer at $30\deg$, while the base of the wind near the disk surface is characterized by Thomson thick with $\tau \sim 13.5 \gg 1$ as expected.

In {\bf Figure~\ref{fig:wind}} the poloidal configuration of each wind ({\tt wind 1-6}) is shown; color-coded (normalized) density distribution $n(r,\theta)$, density contours (dashed curves), poloidal magnetic field lines (solid curves) and poloidal velocity field (white arrows). LoS of $30\deg$ is given in dashed line for reference. 

As expected, the wind density can vary by many orders of magnitude where it is highest near the disk surface at $Z \sim 0$ (i.e. a reservoir of matter) and lowest near the funnel region close to the symmetry axis towards the inner edge of wind.  
Large-scale wind morphology, characterized by $\theta_4$ and $\theta_A$ (in Table~1), indicates a paraboloidal structure as a generic geometry.  
Considering a fiducial viewing angle of $30\deg$ for canonical Seyfert 1 AGNs, note that the LoS does not intersect over a wide range of distance with winds of large opening angle; i.e. wind 4-6,  
allowing X-rays to transmit through wind media without causing obscuration at all. While wind structure can be diverse depending on those variables listed in Table~1, 
as we are interested in disk winds as obscurers in this paper, we further focus only on {\tt wind 1-3} below. 

Note that our wind model (in F10) is steady-state and hence we don't follow its time evolution \citep{BP82,CL94}. Subsequent calculations are made as ``snapshot" at each state, given that the typical timescale of X-ray obscuration spans over many years to a decade, much longer than dynamical timescale for accretion and outflows (unlike local stochastic, turbulent variabilities in the gas). Hence, the wind medium is assumed to reach a quasi-asymptotic, semi-equilibrium state for a given wind morphology with an observer  situated at infinity.

\begin{figure}[t]
\begin{center}
\includegraphics[trim=0in 0in 0in
0in,keepaspectratio=false,width=3.5in,angle=-0,clip=false]{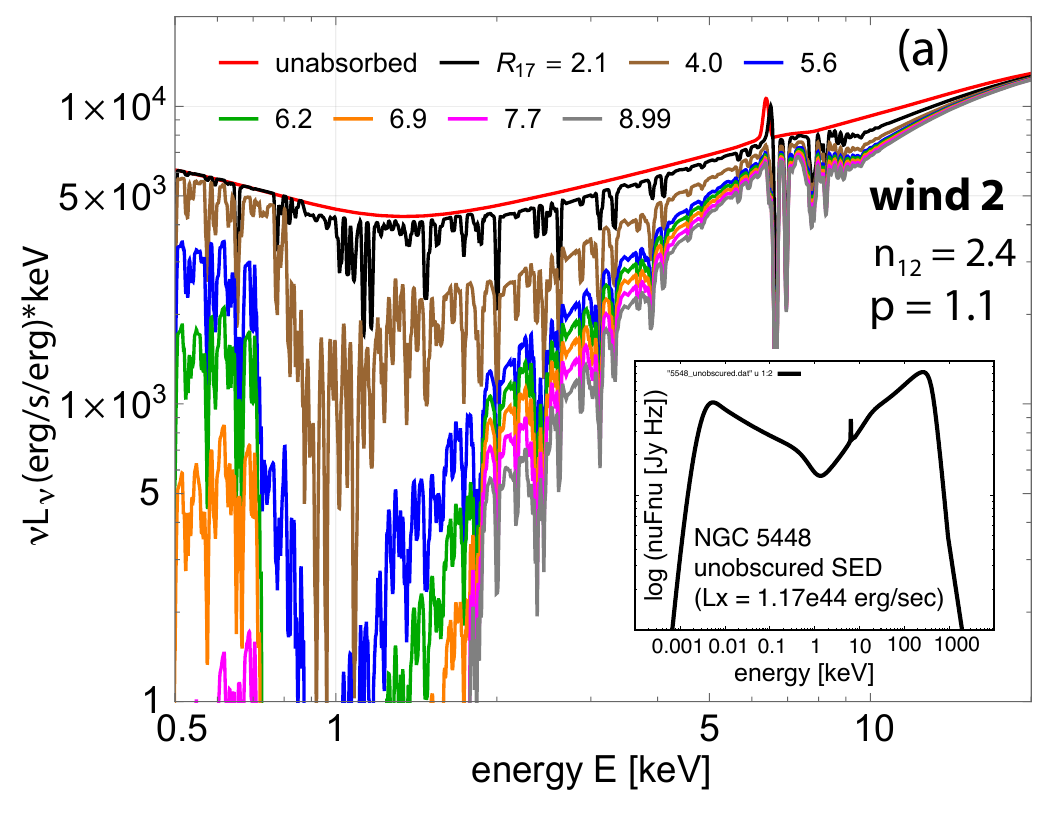}
\includegraphics[trim=0in 0in 0in
0in,keepaspectratio=false,width=3.4in,angle=-0,clip=false]{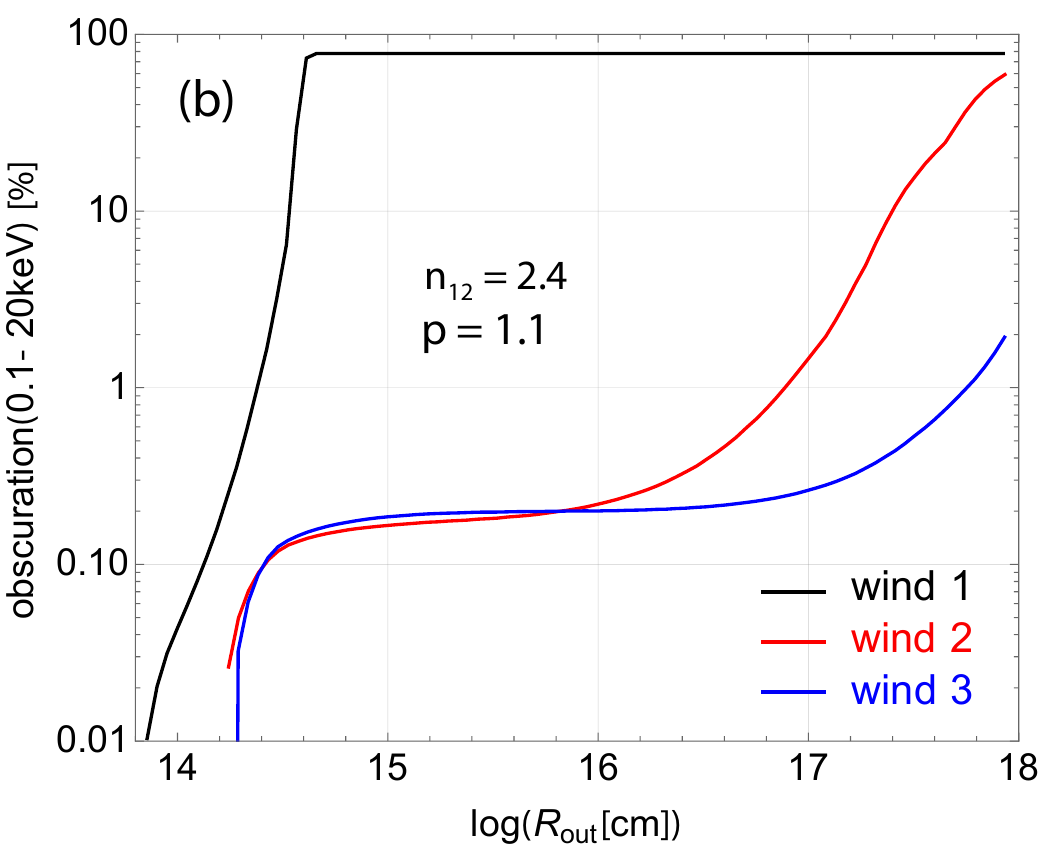}
%
\end{center}
\caption{(a) Calculations of obscured X-ray spectra for ({\tt wind 2}) at different outer extent of the wind $R_{\rm out}$ for $n_{12}=2.4, p=1.1$ and $\theta=30\deg$. Inset in (a) shows the intrinsic (unobscured) SED (in Jy Hz) used for radiative transfer calculations \citep[][]{M15}. (b) Predicted fractional obscuration in 0.1-20 keV flux  relative to the unobscured flux as a function of $R_{\rm out}$ for {\tt wind 1-3} in comparison.     }
\label{fig:obs}
\end{figure}


\section{Modeling of X-Ray Obscuration with Stratified Disk Winds}

Given the physical properties  obtained from {\tt wind 1-3} as discussed in \S 2, radiative transfer calculations are performed with {\tt xstar} with an input (unabsorbed) spectral energy distribution (SED). To this end, we follow the findings from \cite{M15}  for NGC~5548 during its 2013 epoch by employing their reconstructed broadband SED with $L_X = 1.17 \times 10^{44}$ erg~s$^{-1}$ (see inset in {\bf Fig.~\ref{fig:obs}a}). {\bf Figure~\ref{fig:obs}a} shows the predicted X-ray spectra as a function of the outer extent of disk-wind ($r=R_{17} 10^{17}$ cm) for {\tt wind 2} assuming $n_{12}=2.4, p=1.1$ and $\theta=30\deg$. 
In our model, the wind is perfectly smooth and continuous corresponding to a full covering fraction (i.e. $C_{\rm UV,X}=1$) with solar abundances. For $n_{12}=2.4$, the obscuration does not become effective until $r \sim 10^{17}$ cm. 
This is simply because a total cumulative wind column at $r \ll 10^{17}$ cm is not sufficient for causing moderate obscuration of X-rays.
Under this condition, the inner edge of the wind material is highly photoionized at $\log \xi \sim 6.9-7.1$ with gas temperature of $T \sim 10^8$ K. With $p=1.1$, the wind column density falls off very slowly with distance as $d N_H/d(\log \xi)  \propto r^{1-p} \propto \xi^{(p-1)/(2-p)}$ \citep[e.g.][]{K12,F21}. 
%
We find that the magnitude of obscuration is consistent with the wind morphology (most characterized by opening angle); i.e. the more extreme obscuration in {\tt wind 1} and the least case in {\tt wind 3}. As expected, soft X-ray continuum becomes progressively more obscured with increasing wind outer radius $R_{17}$ in all cases. 

The fraction of obscured X-ray flux (0.1-20 keV) relative to the intrinsic (unobscured) one is given in {\bf Figure~\ref{fig:obs}b} as a function of the outer wind radius $R_{\rm out}$ in each case. Wind of smaller opening angle is found to obscure X-ray more effectively at smaller distances; i.e. $r \sim 10^{14-14.5}$ cm in {\tt wind 1} in comparison to $r \gsim 10^{16-18}$ cm in {\tt wind 2} and {\tt wind 3}. 
We find that X-ray is completely blocked  in {\tt wind 1} for a given condition, while only a small fraction of the intrinsic X-ray is absorbed in {\tt wind 3}.   
To further investigate  obscured X-ray spectral shape  relevant for observations, therefore, we follow up on {\tt wind 2}  as a fiducial case. 

\begin{figure}[t]
\begin{center}
\includegraphics[trim=0in 0in 0in
0in,keepaspectratio=false,width=3.1in,angle=-0,clip=false]{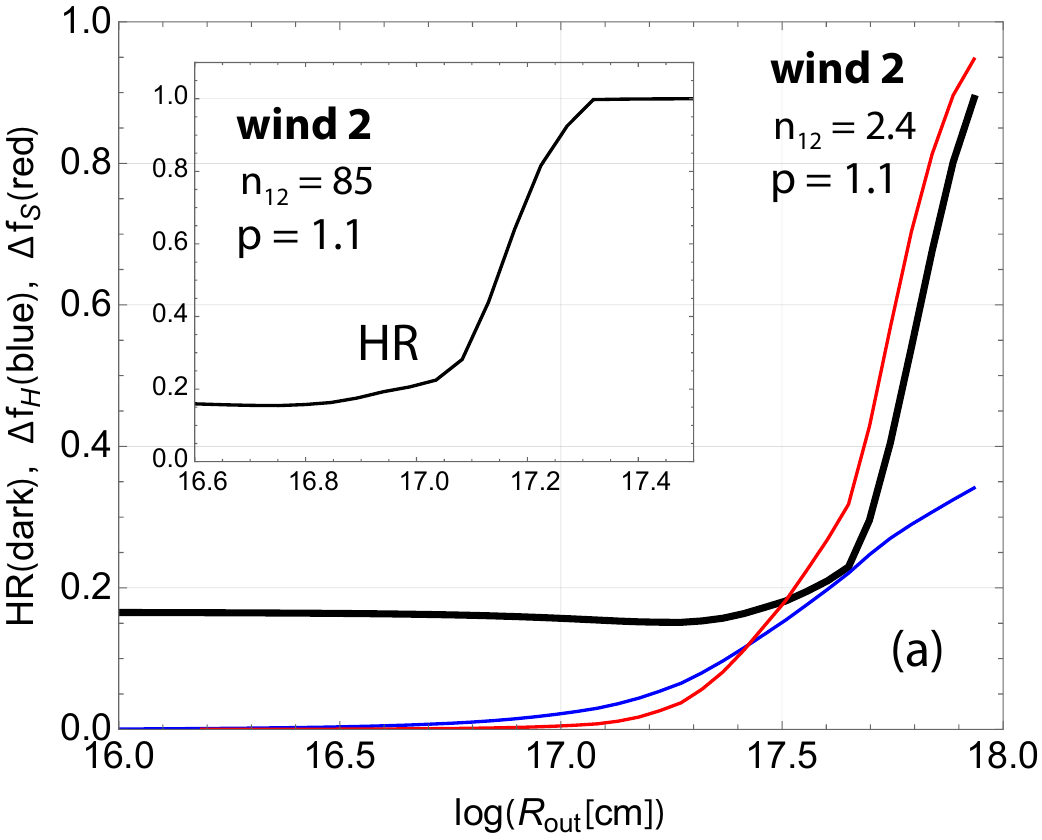}\includegraphics[trim=0in 0in 0in
0in,keepaspectratio=false,width=3.5in,angle=-0,clip=false]{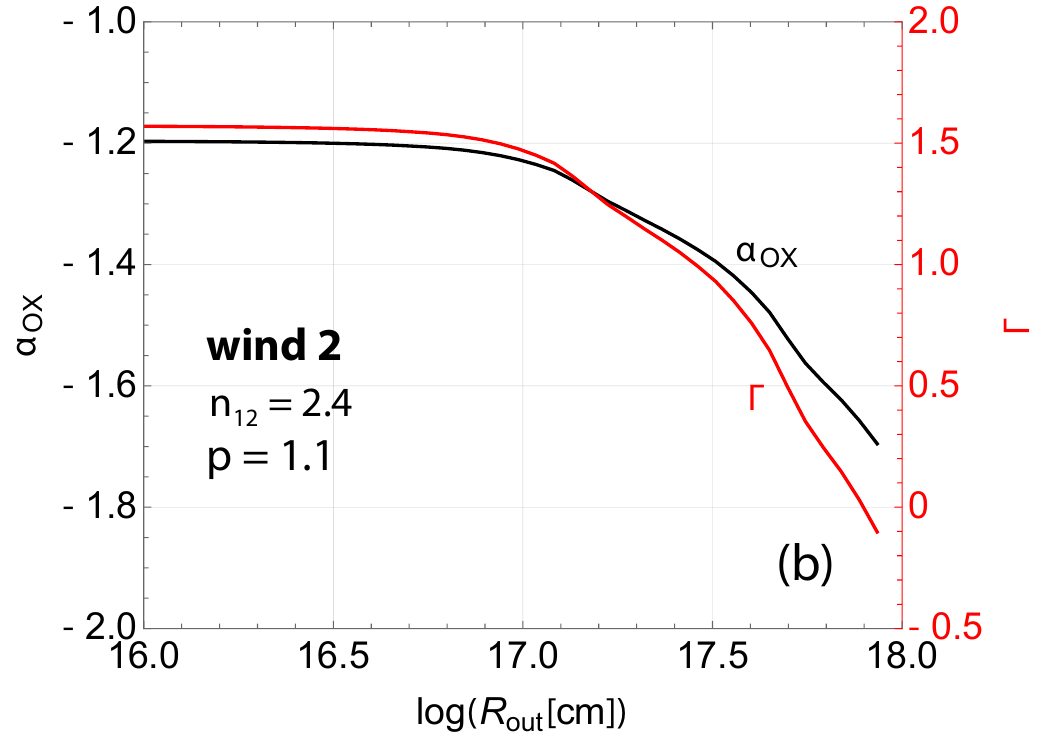}
\end{center}
\caption{(a) Calculated HR of X-ray spectra (dark thick) with the fractional decrease of hard flux $\Delta f_H$ ($2-10$ keV in blue) and soft flux $\Delta f_S$ ($0.2-2$ keV in red) as a function of the outer extent of the wind $R_{\rm out}$  in {\tt wind 2}.  HR with $n_{12}=85$ is shown in inset.  (b) Predicted spectral slope, $\Gamma$ (in red) and $\alpha_{\rm OX}$ (in dark), for $n_{12}=2.4$ in {\tt wind 2}. }
\label{fig:slope}
\end{figure}

In {\bf Figure~\ref{fig:slope}a} we calculate hardness ratio defined as $\rm{HR} \equiv (H-S)/(H+S)$ where $H$ (or $S$) is 2-10 keV (or 0.2-2 keV) flux. 
As expected, higher wind density ($n_{12}=85$) naturally enhances obscuration at smaller distances. From {\bf Figures~\ref{fig:wind}-\ref{fig:slope}a} we learn that X-ray continuum can be very efficiently obscured through the wind, which is also  manifested in a quick rise of HR. This is well understood by the fact that the continuum flux suffers progressively from attenuation by the stratified wind eventually reaching a crucial point where the accumulated column over the LoS becomes substantial. 
%
We also calculate the fractional changes in $H$ by $\Delta f_{H} \equiv |\Delta H|/H_{\rm unobs}$ (in blue) and $S$ by $\Delta f_{S} \equiv |\Delta S|/S_{\rm unobs}$ (in red)  in {\bf Figure~\ref{fig:slope}a} for comparison where ``unobs" denotes unobscured (intrinsic) flux and $\Delta H$ and $\Delta S$ are the flux changes relative to their unobscured values.  
The variation of HR (thus X-ray hardening) is initially caused because both soft ($S$) and hard ($H$) X-ray fluxes are obscured as the wind radius $R_{\rm out}$ increases up to $\sim 10^{17.5}$ cm. As $R_{\rm out}$ extends beyond $\sim 10^{17.5}$ cm,  hard X-ray suppression roughly saturates while soft X-ray continues to get obscured, which results in the change in HR shown in {\bf Figure~\ref{fig:slope}a}.   

We stress here again that the timescale of this change in obscuration is very long (e.g. years; $\sim 9$ years in NGC 5548) and it is reasonable to consider that accretion and wind are in  quasi-equilibrium as $R_{\rm 17}$ varies. These are therefore ``snapshot" calculations. The light-crossing time between $\sim 10^{16}$ cm to $10^{18}$ cm is about one year, although the global wind morphological change (i.e. wind size) should take place at much slower rate.

In parallel, we  examine the broadband spectral shape by computing the optical-to-X-ray strength\footnote[3]{The spectral index $\alpha_{\rm OX} \equiv
0.384 \log (f_{\rm 2keV} / f_{\rm 2500\aa})$ measures the X-ray-to-UV relative brightness where
$f_{\rm 2keV}$ and $f_{\rm 2500\aa}$ are respectively 2 keV and $2500 \aa$
flux densities \citep{Tananbaum79}.}, $\alpha_{\rm OX}$,  in {\tt wind 2} in {\bf Figure~\ref{fig:slope}b} for  $n_{12}=2.4$. As radiation passes through the wind with increasing $R_{\rm out}$, X-ray hardening occurs due to obscuration in soft band (i.e. $\Gamma \sim 1.5 \rightarrow 0$) accompanied by increasing UV counterpart relative to X-ray (i.e. $\alpha_{\rm OX} \sim -1.2 \rightarrow -1.6$) as a consequence.

\section{Predicted X-Ray vs. UV Relations }

\begin{figure}[t]
\begin{center}
\includegraphics[trim=0in 0in 0in
0in,keepaspectratio=false,width=3.5in,angle=-0,clip=false]{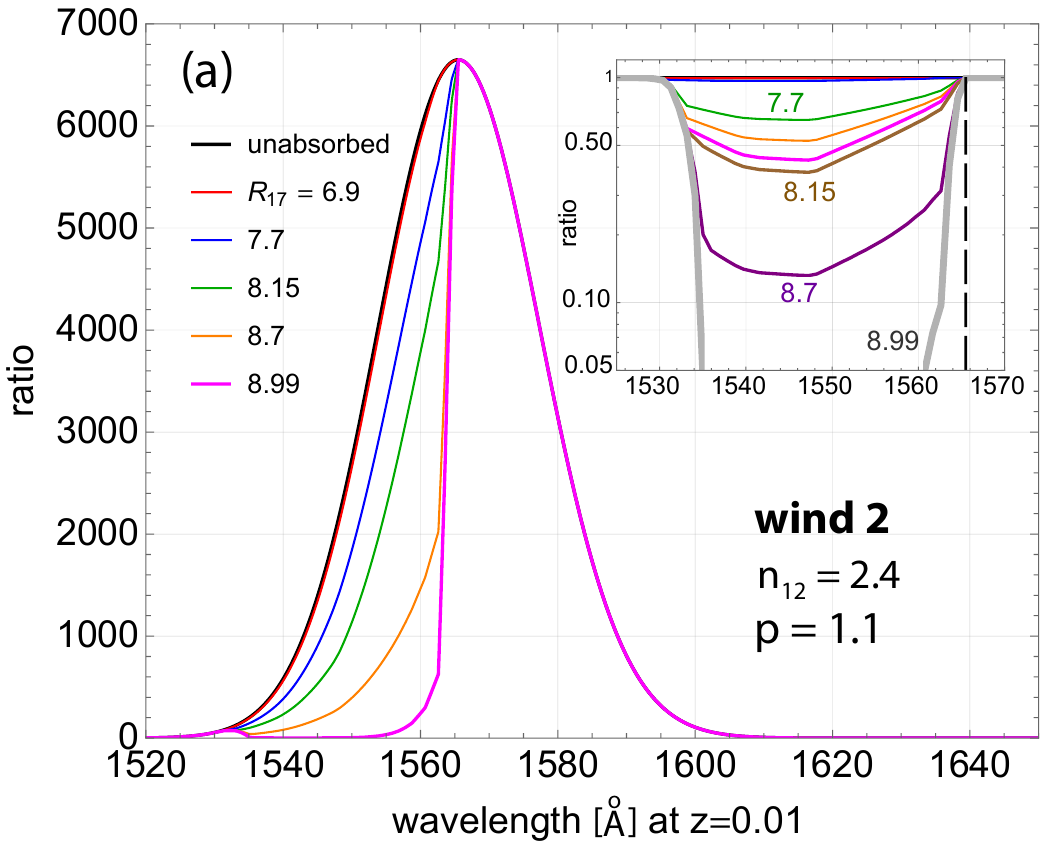}\includegraphics[trim=0in 0in 0in
0in,keepaspectratio=false,width=3.55in,angle=-0,clip=false]{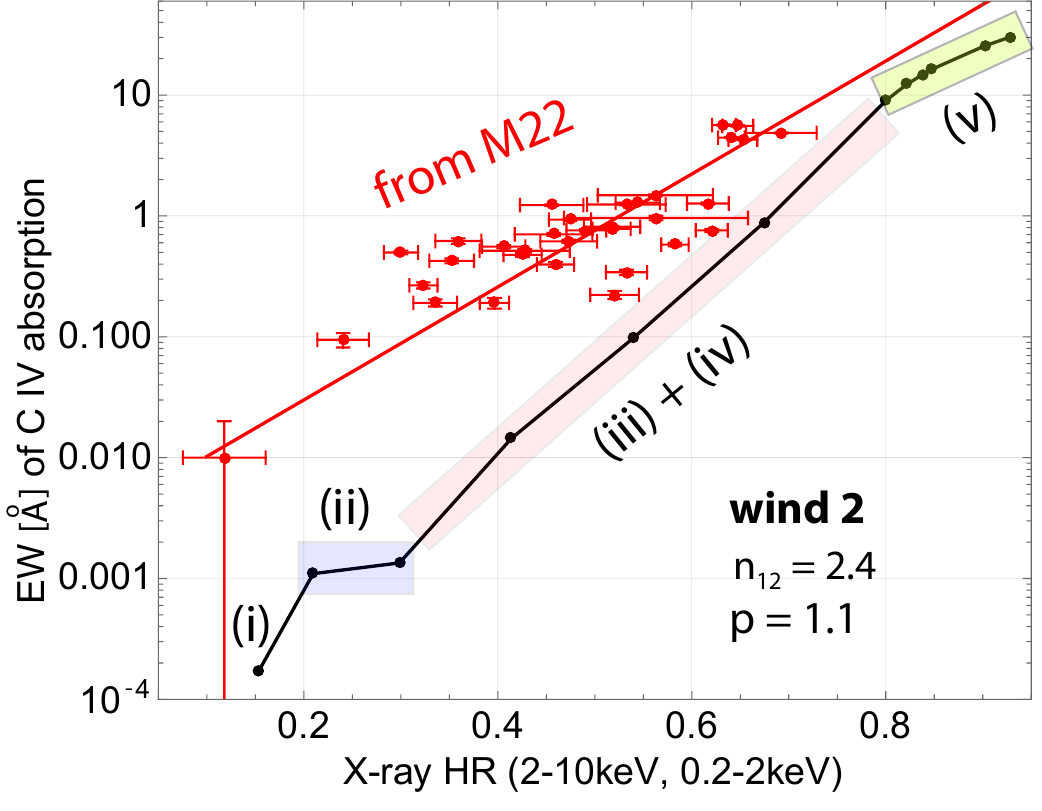}
\end{center}
\caption{(a) Calculations of broad \civ\ absorption imprinted on  Gaussian emission for different outer extent of the wind $R_{17}$ in {\tt wind 2} for $n_{12}=2.4$ 
assuming a source redshift of $z=0.01$ (UV counterpart to {\bf Figure~\ref{fig:obs}a}). Inset shows a progression of \civ\ absorption line with increasing $R_{17}$ (indicated by numbers).  (b) Predicted correlation between X-ray HR and EW (in $\aa$) of \civ\ absorption feature (in dark) in comparison with the observations of NGC~5548 (in red from M22).  }
\label{fig:uvx}
\end{figure}

In relation to X-ray, we consider the transmission of UV continuum due to the same disk winds at larger distances.  
As a diagnostic proxy in UV band  better suited to study BLR, we calculate \civ\ absorption line (a doublet at $1548\aa/1550\aa$) in conjunction with X-ray spectra in {\tt wind 2} for $n_{12}=2.4$, 
by simply assuming that the outer BLR is located within $r \sim 10^{17}$ cm (i.e. $\sim 10^4$ Schwarzschild radii) with the dynamical timescale of $\sim 30$ days or less \citep[e.g.][]{Kaspi05,Homayouni23}. 
In  {\bf Figure~\ref{fig:uvx}a} we show the  signature of broad \civ\ absorption imprinted on  Gaussian \civ\ emission  (assuming a source distance at $z=0.01$)  as a function of the  normalized distance  $R_{17}$ of the outer edge of the wind where where $r=10^{17} R_{17}$ cm. This thus shows UV counterpart to X-ray obscuration given in {\bf Figure~\ref{fig:obs}a}.

An interesting feature is that X-ray is  continuously obscured at distances through $R_{17} \sim 2.1 - 6.9$ ({\bf Fig.~\ref{fig:obs}a}), while absorption feature in \civ\ region is only marginally present. 
This may explain why in some obscured AGNs the observed \civ\ absorption is too weak to be detected; e.g. NGC~3227 \citep[][]{Mao22a} and MR~2251-178 \citep[][]{Mao22b}. 
Beyond $R_{17} \sim 6.9$, the soft X-ray continuum being already  suppressed remains nearly unchanged with increasing distance. On the other hand, the \civ\ absorption is very sensitive to distance; e.g. the absorption continues to become deeper and broader at $R_{17} \lsim 8.15$ with $v_{\rm LoS} \gsim 5,400$ km~s$^{-1}$ and can be almost saturated at $R_{17} \gsim 8.7$ with $v_{\rm LoS} \lsim 4,800$ km~s$^{-1}$, whereas the spectral appearance  of X-ray obscuration seems to vary little. 
This is because the gas of higher column at inner part of the wind can ``self-shield" the outer part of the gas near BLR to lower ionization state for \civ.
Hence, UV counterparts such as \civ\ line can only emerge at distant locations where initially strong ionizing X-rays are  sufficiently weakened by wind media. 
To the best of our knowledge, we are not aware of any  theoretical models trying to account for the observed X-ray obscuration in conjunction with UV absorbers.

The absorption signature becomes less blueshifted (i.e. shifting towards longer wavelength) with increasing $R_{17}$ because of the underlying wind kinematics being $v_{\rm LoS} \propto r^{-1/2}$ as expressed in equation~(1). The feature can get relatively broad suppressing almost the entire blue tail of emission (e.g. $R_{17} \sim 8.7-8.99$). 
%
%
%
These spectral features, X-ray obscuration and UV absorbers, thus appear to be consistent with multi-wavelength observations of NGC~5548 and NGC~3783 (e.g. \citealt{Kaastra14}; \citealt{Kriss19}; M22).


Lastly, we exploit our results in UV/X-ray by extracting a correlation between the equivalent width (EW) of broad \civ\ absorption  and X-ray HR as the outer extent of the wind $R_{17}$ varies. A correlation is clearly seen in {\bf Figure~\ref{fig:uvx}b} where  the following five distinct regimes are identified; 
(i) unobscured state (e.g. $R_{17} \lsim 2.1$) with no/little X-ray obscuration nor UV absorption, (ii) mildly obscured state (e.g. $R_{17} \gsim 2.1 - 4.0$) exhibiting moderate obscuration above $\sim 1$ keV with little/no UV absorption, and (iii) obscured state (e.g. $R_{17} \sim 4.0-5.6$) showing significant obscuration above $\sim 1$ keV with little/no UV absorption and (iv) heavily obscured state ($R_{17} \gsim 5.6-8.0$) even below $\sim 1$ keV together with pronounced  UV absorption signatures, and (v) saturated state ($R_{17} \gsim $8.0) in both UV and X-ray band. 
This trend is partially understood also by the variation of HR shown  in {\bf Figure~\ref{fig:slope}a}. 
As stated in \S 3, attenuation (in X-ray and UV) can be sensitive to the wind size $R_{17}$ because of the stratified structure of continuous density field as  characterized as $n \propto r^{-p}$ (from denser layer to tenuous layer outward in density) where incoming X-rays are progressively absorbed along the LoS. Thus, the larger the wind size is, the more the obscuration is.   
For comparison, the observed correlation from NGC~5548 using HST/COS and Swift  is also shown (in red from M22), which is broadly consistent with the prediction setting aside the exact details (e.g. normalization and slope). 


\section{Summary \& Discussion}

\begin{figure}[t]
\begin{center}
\includegraphics[trim=0in 0in 0in
0in,keepaspectratio=false,width=2.8in,angle=-0,clip=false]{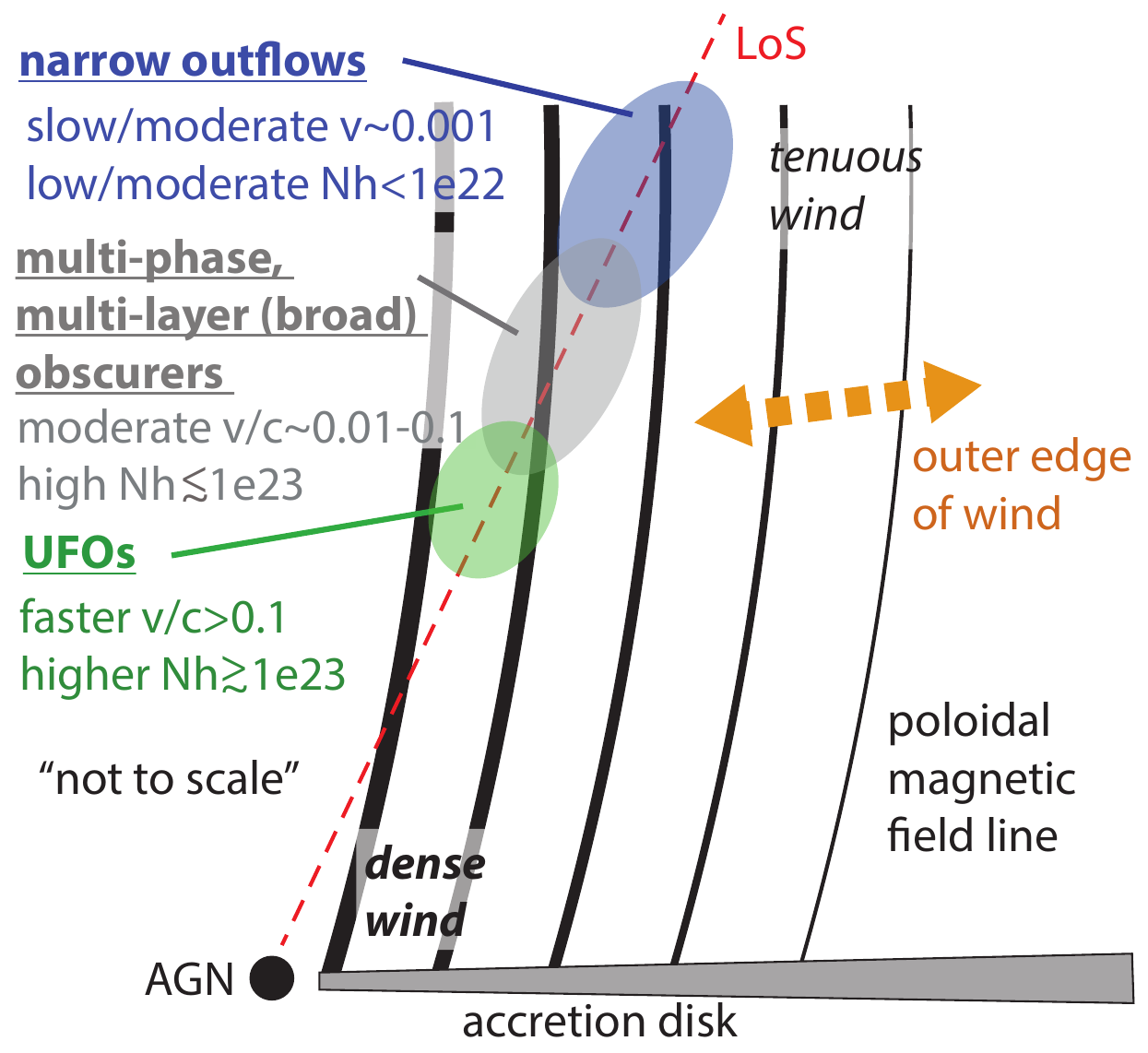}
\end{center}
\caption{Schematic of obscuring accretion disk wind with dual role (UV+X-ray)  in Seyfert 1s in this scenario. {\bf General characteristic  property is indicated.} }
\label{fig:cartoon}
\end{figure}

We consider the dual effects of accretion disk winds of ionized plasma in AGNs  in the context of the variable X-ray obscuration and UV absorption features often observed in  Seyfert 1 galaxies such as NGC~5548 and NGC~3783. As a proof of concept study, we focus on a smooth and continuous winds of stratified structure (as opposed to clumpy winds) by employing an MHD wind launched over a large radial extent of a disk surface. Being coupled to post-process radiative transfer calculations, we compute obscured X-ray spectrum and its UV counterparts. We find that winds of a small opening angle can provide a sufficient obscuring column along a LoS (e.g. $\theta=30\deg$) depending on the density normalization $n_{12}$ at the base of the wind. By simulating a series of broadband spectra as a function of the outer extent of the wind, we demonstrate the following tangible consequences; (1) efficient X-ray obscuration by Compton-thick part of the inner  wind at smaller distances near AGN and (2) UV absorption signatures via \civ\ line attributed to the outer part of the wind around BLR. It is shown that the predicted correlation between X-ray HR and EW of \civ\ absorption line in our model is (at least qualitatively) viable with a number of multi-wavelength data of NGC~5548 (e.g. M22). 
Figure~\ref{fig:cartoon} illustrates the dual role (UV and X-ray) of the continuous disk wind launched over an extended disk surface with a stratified structure in both density and velocity, which might be well represented by MHD winds as considered in this work where the size of the wind (the outer edge) plays a role in characterizing the degree of the dual effect.

In particular, the calculated \civ\ absorption, which can be quite broad depending on the outer extent of the wind, is found to be present exclusively on the blueshifted side of \civ\ emission line profile (see {\bf Figure~\ref{fig:uvx}a}) in  good agreement with the observed HST/STIS spectra from NGC~3783 \citep[e.g.][]{Kriss19} and NGC~5548 (e.g. \citealt{Kaastra14}; M22). The predicted wind velocities of $v_{\rm LoS} \lsim 6,000$ km~s$^{-1}$, responsible for these absorption features, are in consistence with the inferred values from HST/COS data in NGC~5548 \citep[e.g.][]{Kaastra14}.  
We stress that the inner part of the wind helps naturally ``self-shield" (by obscuring soft X-ray photons) the outer part of the same wind from being strongly ionized (to allow UV features at BLR distances) in this framework.

It has been suggested that the observed UV/X-ray phenomena can be well explained by obscuring winds in the form of a clumpy medium through fragmentation perhaps due to thermal instabilities (\citealt{Kaastra14}; \citealt{M15}; \citealt{Kriss19}; M22; \citealt{Waters22}). In this scenario, the behavior of UV/X-ray variabilities (i.e. HR and EW of UV absorption) 
 can be interpreted primarily by variable covering fractions ($C_{\rm UV,X}$), while the obscuring column remains relatively high over multiple epochs; e.g. $C_{\rm UV} \sim 0.01-0.3$ and $C_{\rm X} \sim 0.5-0.95$ while $N_H \sim 10^{23}$ cm$^{-2}$ over a decade in NGC~5548 (M22). As the changes in the observed continuum  and luminosities are only moderate, it  may be thought that the observed variabilities of X-ray continuum and UV absorption are intrinsically caused by a clumpy nature  of obscuring disk winds instead of by varying coronae or disk radiation. 
On the other hand, it is not clear whether those clumpy wind models are in fact capable of  reproducing quantitatively the observed correlation in {\bf Figure~\ref{fig:uvx}b} since no theoretical predictions are made in their framework to date.

Our present work alternatively suggests that the observed UV/X-ray variabilities  can be naturally regulated by the change of disk wind morphology and/or its outer extent (i.e. the outer edge) without needing fragmentation of winds, as demonstrated in {\bf Figure~\ref{fig:uvx}b}. In other words, the outer wind edge (i.e. $R_{\rm out}$) in our work serves a purpose similar to the UV/X-ray covering factors (e.g. $C_{\rm UV,X}$), as illustrated in {\bf Figure~\ref{fig:cartoon}} {\bf where narrow/broad UV/X-ray outflows,  multi-phase, multi-layer obscurers and ultra-fast outflows (UFOs) are partially coexisting at different distances  (see Zaidouni et al.~2024 (submitted) for a similar observational implication in Mrk~817).}

\citet[][]{Partington23} has found that the observed variability in X-ray obscuring column $N_H$ (from {\it NICER}) is closely tied to the change in EW of \siiv\ absorption line from HST/COS in Mrk~817, speculating that the X-ray and UV continua are both impacted by the same obscurers. Our calculations also support this coupling (e.g. obscuring $N_H$ vs. UV line EW), which can be attributed to the change in $R_{\rm out}$  of the stratified disk wind, but not necessarily in a clumpy form.

It is also known that the observed duration of obscuration is different between NGC~3783 ($\sim$ weeks/months) and NGC~5548 ($\sim$ decade), for example. 
Unfortunately, NGC~5548 is the only source so far to provide us with a long-term historical evolution of X-ray obscuration in conjunction with corresponding UV absorption. Furthermore, other variables such as BH mass and Eddington ratio can play a role in influencing the ionization state of obscuring material and its location, regardless of whether it is continuous wind or clumpy wind (or MHD driven or radiation driven). A more detailed understanding of obscuration timescale and duration has to wait until fully dynamical wind simulations can self-consistently integrate multi-dimensional radiative transfer calculations to handle realistic photoionization equilibrium in the gas, which is beyond the scope of the current work.

Such a difference can also be explained by the spatial extent of the wind in the present framework (see also M22).

Our predicted UV/X-ray correlation shown in {\bf Figure~\ref{fig:uvx}b}, while qualitatively consistent, does not quantitatively match the inferred slope and the normalization for NGC~5548 (M22). This discrepancy can be reduced (if not fully resolved) by a  different choice of $n_{12}$ and $p$ in the model.       
As a preliminary study, our results are presented in this work assuming a set of wind parameters mainly represented by  $n_{12}=2.4$ and  $p=1.1$ assuming $\theta=30\deg$. 
We note that the energy range used to compute HR is slightly different between our work and M22, but this does not change the end result.
%
%
In the context of our obscuring wind model, we see that higher value of $n_{12}$ would enhance both obscuration and UV absorption, whereas higher slope of $p$ would lead to a more centrally concentrated wind  near AGN. 
Hence, it is expected that one can reproduce the observed correlation  by,  for example, decreasing $p$-value (e.g. $p \sim 0.9-1.0$) with the same $n_{12}$ value in order to leave sufficient gas at larger distances around BLR to slightly enhance \civ\ EW.   
%
We note that other physical variables such as covering fraction, filling factor, intrinsic change in the continuum can also play some roles in determining the observed correlation, which is beyond the scope of this work. 
A detailed quantitative analysis with multi-wavelength spectroscopy in the future should incorporate these dependences, but it is quite plausible for one to realistically rearrange these variables to better reproduce the derived UV/X-ray correlation.

Historically, Seyfert 2 AGNs are believed to be heavily obscured by Compton-thick gas comprising a ``dusty torus" at $\sim 1-10$ parsec-scale \citep[e.g.][]{Marchesi18,Marchesi22}. 
Its geometrical distribution is still debated, but occultation events and infrared observations favor a ``clumpy torus" viewpoint \citep[e.g.][]{Jaffe04, ElitzurShlosman06}. While not emphasized in this work, it is imaginable that a more permanent obscuration seen in Seyfert 2s, rather than transient one in Seyfert 1s, may well be closely connected to disk-winds of high column considered in this work. A detailed calculation will be needed for further discussion.

In our calculations, we adopt MHD-driven disk wind models (e.g. F10) to demonstrate its viability with data. It is reminded that a pivotal feature of winds needed is a stratified structure (i.e. density and velocity) of a large solid angle over many decades in radius (see {\bf Fig.~\ref{fig:wind}}) and not necessarily restricted to MHD driving alone. While possible, it is never demonstrated explicitly whether winds driven by other equally promising mechanisms (such as thermal driving and radiation driving) may also possess those physical characteristics required for obscuration.

Physics of obscuring disk wind in Seyfert 1s, as either continuous or clumpy media, is yet to be fully explored while providing us with a handful of intriguing insights. For interpreting multi-epoch spectroscopy, it will be very useful to have a long-term monitoring of expected transient and episodic disk winds similar to those available in NGC~5548 and NGC~3783, for example, in an attempt to collectively understand the underlying  variable nature of obscurers. 
A future effort should be able to unambiguously reveal the elusive identity of obscuring winds.

\begin{acknowledgments}
%
The initial idea of this work was inspired through the High Resolution X-Ray Spectroscopy workshop held at MIT in 2023
and we are thankful to an anonymous referee for a number of constructive comments.  
This work is supported in part by NASA/XGS program (proposal: 22-XGS22-0011) through NNH22ZDA001N-XGS.
FT acknowledges funding from the European Union - Next Generation EU, PRIN/MUR 2022 (2022K9N5B4). MD thankfully acknowledges INAF funding under the ``Ricerca Fondamentale 2022" program.
\end{acknowledgments}

\end{document}